\documentclass[manuscript]{aastex}
\usepackage{epsfig,times}
\def\mincir{\raise -2.truept\hbox{\rlap{\hbox{$\sim$}}\raise5.truept \hbox{$<$}\ }}
\def\mincireq{\hbox{\raise0.5ex\hbox{$<\lower1.06ex\hbox{$\kern-1.07em{\sim}$}$}}}
\def\magcir{\raise-2.truept\hbox{\rlap{\hbox{$\sim$}}\raise5.truept \hbox{$>$}\ }}
\def\gr{\kern 2pt\hbox{}^\circ{\kern -2pt K}} 

\def\_{\thinspace}

\shorttitle{BL\,Lacs probe the EBL}
\shortauthors{Mankuzhiyil, Persic \& Tavecchio}

\begin{document}

\title{High-frequency-peaked BL\,Lacertae objects \\ 
	as spectral candles to measure \\ 
	the Extragalactic Background Light \\
	in the Fermi and air Cherenkov telescopes era}

\author{Nijil\,Mankuzhiyil\altaffilmark{1}} 
\affil{Physics Dept., Udine University, via delle Scienze 208, I-33100 Udine UD, Italy}

\author{Massimo\,Persic\altaffilmark{1}} 
\affil{INAF-Trieste, via G.B.\,Tiepolo 11, I-34143 Trieste TS, Italy}

\author{Fabrizio\,Tavecchio} 
\affil{INAF-Brera, via E.\,Bianchi 46, I-23807 Merate LC, Italy}

\altaffiltext{1}{and INFN-Trieste, Gruppo collegato di Udine}


\begin{abstract}
The Extragalactic Background Light (EBL) is the integrated light from all the 
stars that have ever formed, and spans the IR-UV range. The interaction of 
very-high-energy (VHE: $E>100\,$GeV) $\gamma$-rays, emitted by sources located 
at cosmological distances, with the intervening EBL results in $e^-e^+$ pair 
production that leads to energy-dependent attenuation of the observed VHE flux. 
This introduces a fundamental ambiguity in the interpretation of measured 
VHE\,$\gamma$-ray spectra: neither the intrinsic spectrum, nor the EBL, are 
separately known -- only their combination is. 
In this paper we propose a method to measure the EBL photon number density. 
It relies on using simultaneous observations of BL\,Lac objects in the 
optical, X-ray, 
high-energy (HE: $E>100\,$MeV) $\gamma$-ray (from the Fermi telescope), and 
VHE\,$\gamma$-ray (from Cherenkov telescopes) bands. For each source, the method 
involves best-fitting the spectral energy distribution (SED) from optical through 
HE\,$\gamma$-rays (the latter being largely unaffected by EBL attenuation as long 
as $z \mincir 1$) with a Synchrotron Self-Compton (SSC) model. We extrapolate 
such best-fitting models into the VHE regime, and assume they represent the BL\,Lacs'
intrinsic emission. Contrasting measured versus intrinsic emission leads to a 
determination of the $\gamma\gamma$ opacity to VHE photons. 
Using, for 
each given source, different states of emission will only improve the accuracy of 
the proposed method. We demonstrate this method using recent simultaneous 
multi-frequency observations of the 
high-frequency-peaked BL\,Lac object PKS\,2155-304 and discuss how similar 
observations can more accurately probe the EBL.  
\end{abstract}

\keywords{ 
BL Lacertae objects: general -- 
BL Lacertae objects: individual (PKS 2155-304) --
diffuse radiation --
gamma rays: galaxies --
infrared: diffuse background 
}

\section{Introduction}

The Extragalactic Background Light (EBL), in both its level and degree of cosmic 
evolution, reflects the time integrated history of light production and re-processing 
in the Universe, hence the history of cosmological star-formation. Roughly speaking, 
its shape must reflect the two humps that characterize the spectral energy distributions 
(SEDs) of galaxies: one arising from starlight and peaking at $\lambda \sim 1\,\mu$m 
(optical background), and one arising from warm dust emission and peaking at $\lambda 
\sim 100\,\mu$m (infrared background). However, direct measurements of the EBL are 
hampered by the dominance of foreground emission (interplanetary dust and Galactic 
emission), hence the level of EBL emission is uncertain by a factor of several. 

One approach to evaluate the EBL emission level has been modeling the integrated light 
that arises from an evolving population of galactic stellar populations. However, 
uncertainties in the assumed galaxy formation and evolution scenarios, stellar initial 
mass function, and star formation rate have led to significant discrepancy among models 
(e.g., Salamon \& Stecker 1998; Malkan \& Stecker 1998 and Stecker \& de\,Jager 1998; 
Kneiske et al. 2002 and 2004; Stecker, Malkan \& Scully 2006; Razzaque, Dermer \& Finke 
2009 and Finke, Razzaque \& Dermer 2010). These models have been used to correct observed 
VHE spectra and deduce (EBL model dependent) 'intrinsic' VHE\,$\gamma$-ray emissions. 

The opposite approach, of a more phenomenological kind, deduces upper limits on the level 
of EBL attenuation making basic assumptions on the intrinsic VHE\,$\gamma$-ray shape of 
AGN spectra. Specifically, it was assumed that the latter are described by a power-law 
photon index $\Gamma \geq 1.8$ (Schroedter 2005), $\Gamma \geq 1.5$ (e.g., Aharonian et al. 
2006; Mazin \& Goebel 2007; Mazin \& Raue 2007), and $\Gamma \geq 1$ (Finke \& Razzaque 
2009). These assumptions correspond to various possibilities of producing TeV spectra. 
Shock-accelerated electrons are unikely to produce VHE\,$\gamma$-rays with $\Gamma < 1.5$ 
from Compton scattering (e.g., Blandford \& Eichler 1987). However, either internal $\gamma 
\gamma$ absorption (Aharonian et al. 2008), or harder electron spectra at the highest 
energies in relativistic shocks (Stecker, Baring \& Summerlin 2007), or Compton scattering 
of CMB photons (B\"ottcher et al. 2008), or top-heavy power-law energy distributions of the 
emitting electrons (Katarzynski et al. 2006) could lead to harder intrinsic TeV spectra -- 
not to mention that pion decay from a hadronic source would produce a very hard TeV component, 
irrespective of the lower-energy electron synchrotron spectrum (M\"ucke et al. 2003). Another 
proposed approach to derive EBL upper limits involves assuming that a same-slope extrapolation 
of the observed {\it Fermi}/LAT HE spectrum into the VHE domain exceeds the intrinsic VHE 
spectrum there (Georganopulos, Finke \& Reyes 2010). An approach to exploring the redshift 
evolution of the EBL exploits the GeV-TeV connection for blazar spectra (Stecker \& Scully 2010).

A different, but related, approach to constraining the EBL rests on evaluating the 
collective blazar contribution to the extragalactic $\gamma$-ray background when the 
inescapable electromagnetic cascades of lower-energy photons and electrons, initiated 
by the interaction of VHE photons with the EBL, are accounted for. The collective 
intensity of a cosmological population of VHE\,$\gamma$-ray sources will be attenuated 
at the highest energies through interaction with the EBL and enhanced at lower energies 
by the resulting cascade: the strength of the effect depends on the source $\gamma$-ray 
luminosity function and spectral index distribution, and on the EBL model (Venters 2010). 
The extragalactic $\gamma$-ray background, resulting from the contributions of different 
classes of blazars, can then be used to constrain the EBL (Kneiske \& Mannheim 2008; 
Venters, Pavlidou \& Reyes 2009) -- even though the amount of energy flux absorbed and 
reprocessed is probably only a small fraction of the total extragalactic $\gamma$-ray 
background energy flux (Inoue \& Totani 2009). 

The only unquestionable constraints on the EBL are model-independent lower limits 
based on galaxy counts (Dole et al. 2006; Franceschini, Rodighiero \& Vaccari 2008). 
It should be noted, however, that the EBL upper limits in the 2--80$\mu$m obtained 
by Mazin \& Raue (2007) combining results from all known TeV blazar spectra (based 
on the assumption that the intrinsic $\Gamma \geq 1.5$) are only a factor 
$\approx$2--2.5 above the absolute lower limits from source counts. So it would 
appear that there is little room for additional components like Pop\,III stars (Raue, 
Kneiske \& Mazin 2009; Aharonian et al. 2006), unless we miss some fundamental aspects 
of blazar emission theory (which have not been observed in local sources, however).

An attempt to measure the EBL used the relatively faraway blazar 3C\,279 as a 
background light source (Stecker, de\,Jager \& Salamon 1992), assuming that the 
intrinsic VHE spectrum was known from extrapolating an apparently perfect $E^{-2}$ 
power-law differential energy spectrum, known in the interval from 70\,MeV to $>$5\,GeV 
from EGRET data, by a couple of further decades in energy into the VHE regime. However, 
blazars are highly variable sources, so it's almost impossible to determine with 
confidence the intrinsic TeV spectrum -- which itself can be variable. 

In this paper we propose a method to measure the EBL that improves on Stecker et al. 
(1992) by making a more realistic assumption on the intrinsic TeV spectrum. Simultaneous 
optical/X-ray/HE/VHE (i.e., eV/keV/GeV/TeV) data are crucial to this method, considering 
the strong and rapid variability displayed by most blazars. After reviewing features of 
EBL absorption (sect.\,2) and of the adopted BL\,Lac emission model (sect.\,3), in 
sect.\,4 we describe our technique, in sect.\,5 we apply it to recent multifrequency 
observations of PKS\,2155-304 and determine the $\gamma \gamma$ optical depth out to that 
source's redshift. In sect.\,6 we discuss our results.

\section{EBL absorption}

The cross section for the reaction $\gamma \gamma \rightarrow {\rm e}^+{\rm e}^-$ is 
\begin{eqnarray}
\lefteqn{
\sigma_{\gamma\gamma}(E, \epsilon, \phi) ~=~ {3 \over 16}\, \sigma_{\rm T} ~(1-\beta^2)~~\times }
	\nonumber\\
 & & 
\times ~~ \biggl[\, 2\,\beta\,(\beta^2-2) ~+~ 
(3-\beta^4) ~{\rm ln} {1+\beta \over 1-\beta} \, \biggr]\, 
\end{eqnarray}
(see Stecker et al. 1992), where $\sigma_{\rm T}$ is the Thompson cross section and 
$\beta(E, \epsilon, \phi) \equiv \sqrt{1 - 2(m_ec^2)^2 /E \epsilon (1- {\rm cos}\,\phi) }$ with 
$\phi$ the angle between the photons of energy, respectively, $E$ ('hard') and $\epsilon$ ('soft'). 

Purely for analytical demonstration purposes we assume, following Stecker et al. (1992), that 
$n_{\rm EBL}(\epsilon) \propto \epsilon^{-2.55}$ is the local number density of EBL photons 
having energy equal to $\epsilon$ as appropriate in the mid- to far-infrared ($\sim$20-100 $\mu$m) 
range 
\footnote{
	The EBL has a two-bump spectral shape, and does not follow a 
      simple power law but a polynomial of higher order.
}
(no redshift evolution -- as befits the relatively low redshifts currently accessible to Cherenkov 
telescopes), $z_{\rm s}$ is the source redshift, and the cosmology is flat no-$\Lambda$ ($\Omega_0=1$)
\footnote{ 
	The main result does not change if the currently favored concordance cosmology is used.}. 
The optical depth due to pair-creation attenuation between the source and the Earth, 
\begin{eqnarray}
\lefteqn{
\tau_{\gamma \gamma}(E, z_{\rm s}) ~=~ {c \over H_0} \int_0^{z_{\rm s}} \sqrt{1+z}~ {\rm d}z ~ 
\int_0^2 {x \over 2} {\rm d}x ~~~ \times} 
                \nonumber\\
 & & 
\times ~ \int_{2(m_ec^2)^2 \over Ex(1+z)^2}^\infty n_{\rm EBL}(\epsilon)~~ \sigma_{\gamma 
\gamma}\bigl(2xE \epsilon (1+z)^2\bigr) ~~ {\rm d}\epsilon
\end{eqnarray}
where $x \equiv (1 - {\rm cos}\,\phi)$ and $H_0$ being the Hubble constant, turns out to be 
$\tau_{\gamma \gamma}(E,z) \propto E^{1.55} z_{\rm s}^{\eta}$ with $\eta \sim 1.5$. 

\begin{figure*}

\vspace{6.5cm}
\includegraphics{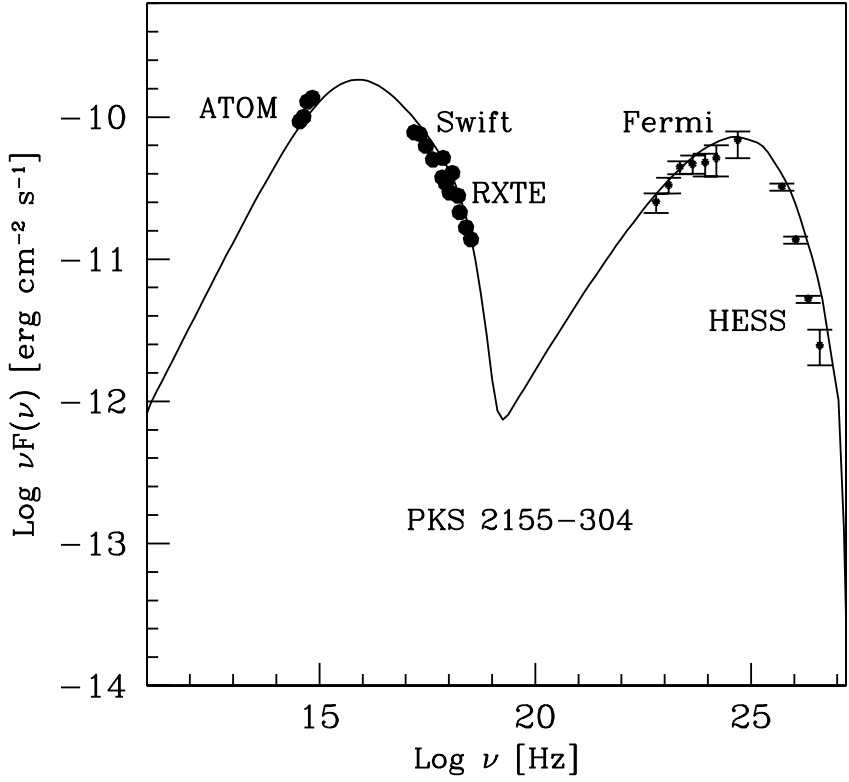}

\caption{ 
Data (symbols: from Aharonian et al. 2009; no VHE\,$\gamma$-ray upper limits reported) 
and best-fit one-zone SSC model (solid curve) of the SED of PKS\,2155-304. 
The best-fit ($\chi^2_\nu = 0.78$ for $\nu = 16$ degrees of freedom) SSC parameters are:  
$n_{\rm e}=150$\,cm$^{-3}$, 
$\gamma_{\rm br}=2.9 \times 10^4$,  
$\gamma_{\rm max}=8 \times 10^5$,
$\alpha_1=1.8$,
$\alpha_2=3.8$,
$R=3.87 \times 10^{16}$\,cm, 
$\delta=29.2$, 
$B=0.056$\,G.
The obtained values of $R$ and $\delta$ imply a variability 
timescale $t_{\rm var} \sim R\,(1+z) /(c \delta)$, which is 
compatible with the observed value of $\approx$12\,hr. 
}
\end{figure*}

This calculation, although it refers to an idealized case, highlights an important 
property of the VHE flux attenuation by the $\gamma_{\rm VHE} \gamma_{\rm EBL} \rightarrow 
{\rm e}^+{\rm e}^-$ interaction: $\tau_{\gamma\gamma}$ depends both on the distance 
traveled by the VHE photon (hence on $z$) and on the photon's (measured) energy $E$.
So the spectrum measured at Earth is distorted with respect to the emitted spectrum.
In detail, the expected VHE $\gamma$-ray flux at Earth will be: $F(E)$$=($d$I/$d$E)\,e^
{-\tau_{\gamma \gamma}(E)}$ (differential) and $F($$>$$E)$$=$$\int_E^\infty ($d$I/$d$E^
\prime)\,e^{-\tau_{\gamma \gamma}(E^{\prime})}dE^\prime$ (integral).

\section{BL\,Lac SSC emission}

In order to reduce the degrees of freedom, we use a simple one-zone SSC model (for 
details see Tavecchio, Maraschi \& Ghisellini 1998, and Tavecchio \& Maraschi 2003). 
This has been shown to adequately describe broad-band SEDs of most high-frequency-peaked 
BL\,Lac objects (HBLs; e.g., Ghisellini et al. 1998) and, for a given source, both its 
ground and excited states (Tavecchio et al. 2001; Tagliaferri et al. 2008). The main 
support for the one-zone model is that in most such sources the temporal variability is 
clearly dominated by one characteristic timescale, which implies one dominant characteristic 
size of the emitting region (e.g., Anderhub et al.  2009). Moreover, one of the most 
convincing evidence favouring the SSC model is the strict correlation between the variations 
in the X-ray and in the TeV band (e.g., Fossati et al. 2008). Since in the SSC model the 
emission in the two bands is produced by the same electrons (via synchrotron and SSC 
mechanism, respectively), a strict correlation is expected. However, there are (rare) 
exceptions, e.g. the so-called ``orphan'' TeV flares (Krawczynski et al. 2004) which 
are not accompanied by corresponding variations in the X-ray band; or more complex models 
than the simple one-zone SSC model may be required to fit the observed variability pattern 
in some cases (e.g., the July 2006 flare of PKS\,2155-304: see Costamante 2008; but also 
Foschini et al. 2007 and Kusunose \& Takahara 2008).

The emission zone is supposed to be spherical with radius $R$, in relativistic 
motion with bulk 
Lorentz factor $\Gamma$ at an angle $\theta $ with respect to the line of sight 
to the observer, so that 
special relativistic effects are cumulatively described by the relativistic Doppler factor, 
$\delta=[\Gamma(1-\beta\,{\rm cos}\,\theta)]^{-1}$. Relativistic electrons with 
density $n_{\rm e}$ and a
tangled magnetic field with intensity $B$ homogeneously fill the region. The 
relativistic electrons' spectrum is described by a smoothed broken 
power-law function of the electron Lorentz factor $\gamma$, with limits $\gamma _1$ 
and $\gamma _2$ and break at $\gamma _{\rm br}$ and low- and high-energy slopes 
$\alpha_1$ and $\alpha_2$. This purely phenomenological choice 
is motivated by the observed shape of the bumps in the SEDs, well represented by two 
smoothly joined power laws. In calculating the SSC emission we use the full 
Klein-Nishina cross section, especially important in shaping the TeV spectrum. 

As detailed in Tavecchio et al. (1998), this simple model can be fully constrained 
by using simultaneous multifrequency observations. Indeed, the model's free 
parameters are 9, of which 6 specify the electron 
energy distribution ($n_{\rm e}$, $\gamma_1$, $\gamma_{\rm br}$, $\gamma_2$, $\alpha_1$, 
$\alpha_2$), and 3 describe the global properties of the emitting region ($B$, $R$, $\delta$). 
On the other hand, from observations ideally 
one can derive 9 observational quantities: the slopes of the synchrotron bump after 
and above the peak $\alpha _{1,2}$ (uniquely connected to $n_{1,2}$), the synchrotron 
and SSC peak frequencies ($\nu _{\rm s,C}$) and luminosities $L_{\rm s,C}$, and the 
minimum variability timescale $t_{\rm var}$ which provides an estimate of the size of 
the sources through $R < c t_{\rm var} \delta /(1+z)$. 

Therefore, once the relevant observational quantities are known, one can uniquely 
derive the set of SSC parameters.

\section{The method}

The method we are proposing stems from the consideration that both the 
EBL and the intrinsic VHE\,$\gamma$-ray spectra of background sources 
are fundamentally unknown. In order to measure the EBL at different 
$z$, one should single out a class of sources that is homogeneous, i.e. 
it can be described by one same emission model at all redshifts. This 
approach is meant to minimize biases that may possibly arise from 
systematically different SED modelings adopted for different classes 
of sources at different distances. So we choose the class of source 
that has both a relatively simple emission model and the potential of 
being seen from large distances: BL\,Lac objects, i.e. AGNs whose relativistic 
jets point directly toward the observer so their luminosities are boosted by a 
large factor and dominate the source flux with their SSC emission. Within 
BL\,Lacs, we propose to use the sub-class of HBL, because their Compton peak 
can be more readily detected by 
Cherenkov telescopes than other types of source, and because their HE 
spectrum can be described as a single (unbroken) power law in photon 
energy, unlikely other types of BL\,Lacs (Abdo et al. 2009). 

For a given BL\,Lac, our method relies on using, a simultaneous broad-band 
SED that samples the optical, X-ray, high-energy (HE: $E>100\,$MeV) 
$\gamma$-ray (from the {\it Fermi} telescope), and VHE\,$\gamma$-ray 
(from Cherenkov telescopes) bands. A given SED will be best-fitted, 
from optical through HE\,$\gamma$-rays, with a Synchrotron Self-Compton 
(SSC) model. [Photons with $E \mincir 100$\,GeV are largely unaffected 
by EBL attenuation (for reasonable EBL models) as long as $z \mincir 1$.] 
Extrapolating such best-fitting SED model into the VHE regime, we shall 
assume it represents the source's intrinsic emission. It should be 
emphasized that the electrons that are responsible for such ''intrinsic'' 
VHE emission are the same that have been simultaneously and cospatially 
measured through their synchrotron emission in the optical and X-ray bands 
and the Compton emission in the HE\,$\gamma$-ray band. As emphasized by Coppi 
\& Aharonian (1999), only if the X-ray/$\gamma$-ray variations are consistent 
with being produced by a common electron ditribution, then it is possible to 
robustly estimate a BL\,Lac's intrinsic TeV spectrum from its emission at 
lower energies. 
\footnote{
	However, the electrons resulting from the interactions of the 
	VHE photons and the EBL can upscatter EBL and CMB photons along 
	the line of sight from the source to the observer, and these 
	would also contribute to the observed SED (e.g., Venters, Pavlidou 
	\& Reyes 2009; Venters 2010; Inoue \& Totani 2009; Kneiske \& 
	Mannheim 2008): if this contribution is significant, it could 
	complicate the connections between the observations of sources 
	in the various energy bands.
} 
Contrasting measured 
versus intrinsic emission yields a determination of exp$[-\tau_{\gamma\gamma}
(E,\,z)]$, the energy-dependent absorption of the VHE emission coming from a 
source located at redshift $z$ due to pair production with intervening EBL 
photons. Once $\tau_{\gamma\gamma}(E,\,z)$ is known, by assuming a 
specific cosmology one can derive the EBL photon number density: e.g., from 
$\tau_{\gamma\gamma}(E,\,z) = \int \int \int \sigma_{\gamma\gamma}(E,\,\epsilon, 
\theta)\, n_{\rm EBL}(\epsilon) \,{\rm d}\theta \,{\rm d}\epsilon \,{\rm d}\ell$ 
(see Eq.\,2), if we have $k$ values of $\tau_{\gamma \gamma}$ for $k$ different 
values of $E$, by adopting a parametric form of $n_{\rm EBL}(\epsilon) = 
\sum_{\rm 0\,j}^{\rm k} a_{\rm j}\epsilon^{\rm j}$ in principle we can solve 
for the $k$ coefficients $a_{\rm j}$.

Using SEDs from different HBLs and, for a given source, from different 
states of emission, will improve the accuracy of the method by increasing the 
number of EBL measurements. 

Setting $n_{\rm EBL}(\epsilon,z) = n_{\rm EBL}(\epsilon) (1+z)^{\kappa}$, 
repeating the procedure at different redshift shells would allow us to estimate 
the EBL cosmic evolution rate parameter $\kappa$. 

Clearly this approach could only have been used starting from the current epoch, 
because of the availability of {\it Fermi}/LAT data simultaneous with optical and 
X-ray data that allow us to substantially remove the degeneracy of the SSC model 
at low energies and hence to estimate, for the first time, the intrinsic TeV emission. 
The concomitant availability of simultaneous air Cherenkov data enables a 
measurement of the EBL opacity -- albeit in a model-dependent way. We here propose 
using a homogeneous sample of same-emission sources to derive a coherent, unbiased 
picture of the EBL.

\subsection{Best-fit procedure: $\chi^2$ minimization}

In order to fit the observed optical, X-ray and HE $\gamma$-ray flux with the SSC 
model, a $\chi^{2}$ minimization is used. We vary the SSC model's 9 parameters by 
small logarithmic steps. If the variability timescale of the flux, $t_{\rm var}$, 
is known, one can set $R \sim c t_{\rm var} \delta /(1+z)$, so the free parameters are 
reduced to 8. We assume here $\gamma_{\rm min} = 1$: for a plasma with $n_{\rm e} 
\approx {\cal O}(10)$\,cm$^{-3}$ and $B \approx {\cal O}$(0.1)\,G (as generally 
appropriate for TeV BL\,Lac jets: e.g., Ghisellini et al. 1998; Costamante 
\& Ghisellini 2002; Finke, Dermer \& B\"ottcher 2008), this approximately corresponds 
to the energy below which Coulomb losses exceed the synchrotron losses (e.g., Rephaeli 
1979) and hence the electron spectrum bends over and no longer is power-law. However, in 
general $\gamma_{\rm min}$ should be left to vary -- e.g., cases of a "narrow" 
Compton component require $\gamma_{\rm min}$$>$1 (Tavecchio et al. 2010). 

In order to reduce the run time of the code, the steps are adjusted in each run 
such that, a larger $\chi^2$ is followed by larger steps.

\section{Example: application to PKS\,2155-304}

We apply the procedure described in Sect.\,4 to the simultaneous SED data set 
of PKS\,2155-304 described in Aharonian et al. (2009). The data and resulting best-fit 
SSC model (from optical through HE\,$\gamma$-rays) are shown in Fig.(1). 
The extrapolation of the model into the VHE\,$\gamma$-ray range clearly lies below 
the observational H.E.S.S. data, progressively so with increasing energy. We 
attribute this effect to EBL attenuation, $F_{\rm obs}(E;\,z)$$=$$F_{\rm 
em}(E;\,z)\, e^{-\tau_{\gamma\gamma}(E;\,z)}$. The corresponding values of 
$\tau_{\gamma\gamma}(E;\,z)$ for $E$$=$0.23, 0.44, 0.88, 1.70\,TeV and a source 
redshift $z$$=$0.116 are, respectively, $\tau_{\gamma\gamma}=0.12$, 0.48, 0.80, and 0.87 . 

We note that the SED analysis of Aharonian et al. (2009) was based on a slightly 
different SSC model, that involved a three-slope (as opposed to our two-slope) 
electron spectrum. This difference may lead to a somewhat different decreasing 
wing of the modeled Compton hump, and hence to a systematic difference in the 
derived $\tau_{\gamma\gamma}(E;\,z)$. That said, it's however interesting to note 
that the main parameters describing the plasma blob ($B$, $\delta$, $n_{\rm e}$) 
take on similar values in our best-fit analysis and in Aharonian et al. (2009). 
More generally, these parameters are quite similar to those deduced for other 
$\gamma$-ray BL\,Lac (see Tavecchio et al. 2010). 

In Fig.(2) we compare our determination of $\tau_{\gamma\gamma}$ with 
some recent results (Franceschini et al. 2008; Gilmore et al. 2009; Kneiske et al. 
2004; Stecker et al. 2006; Finke et al. 2010; Raue \& Mazin 2008). Whereas 
our values are generally compatible with previously published constraints, we note 
that our values -- that refer to a source redshift $z=0.116$ -- closely agree 
with the corresponding values of Franceschini et al. (2008; see their curve 
corresponding to $z=0.10$), which are derived from galaxy number counts and hence 
represent the light contributed by the stellar populations of galaxies prior to the 
epoch corresponding to source redshift $z_{\rm s}$ -- i.e., the minimum amount (i.e., 
the guaranteed level) of EBL (see also Malkan \& Stecker 2001). 

\begin{figure*}

\vspace{6.5cm}
\includegraphics{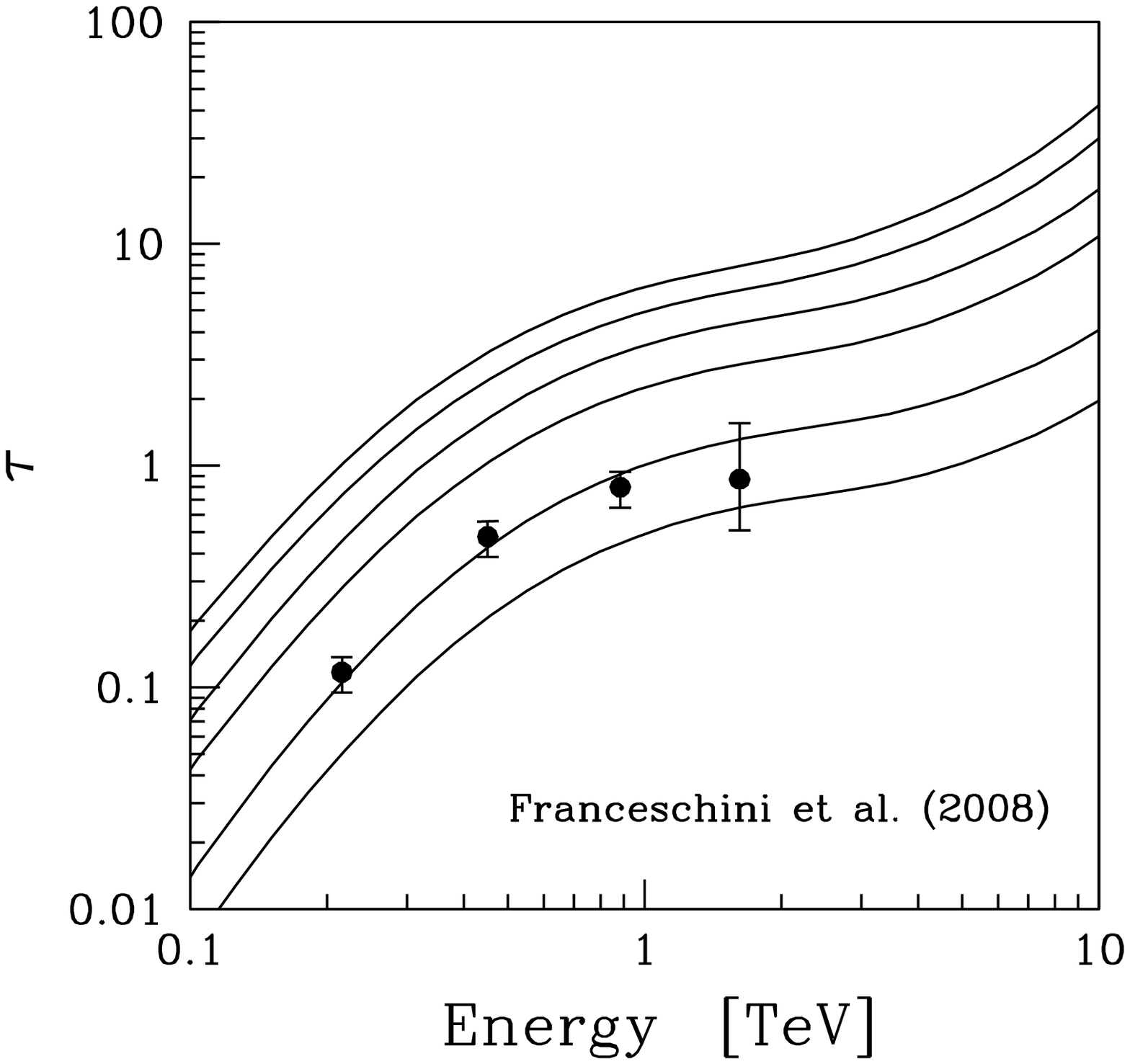}
\includegraphics{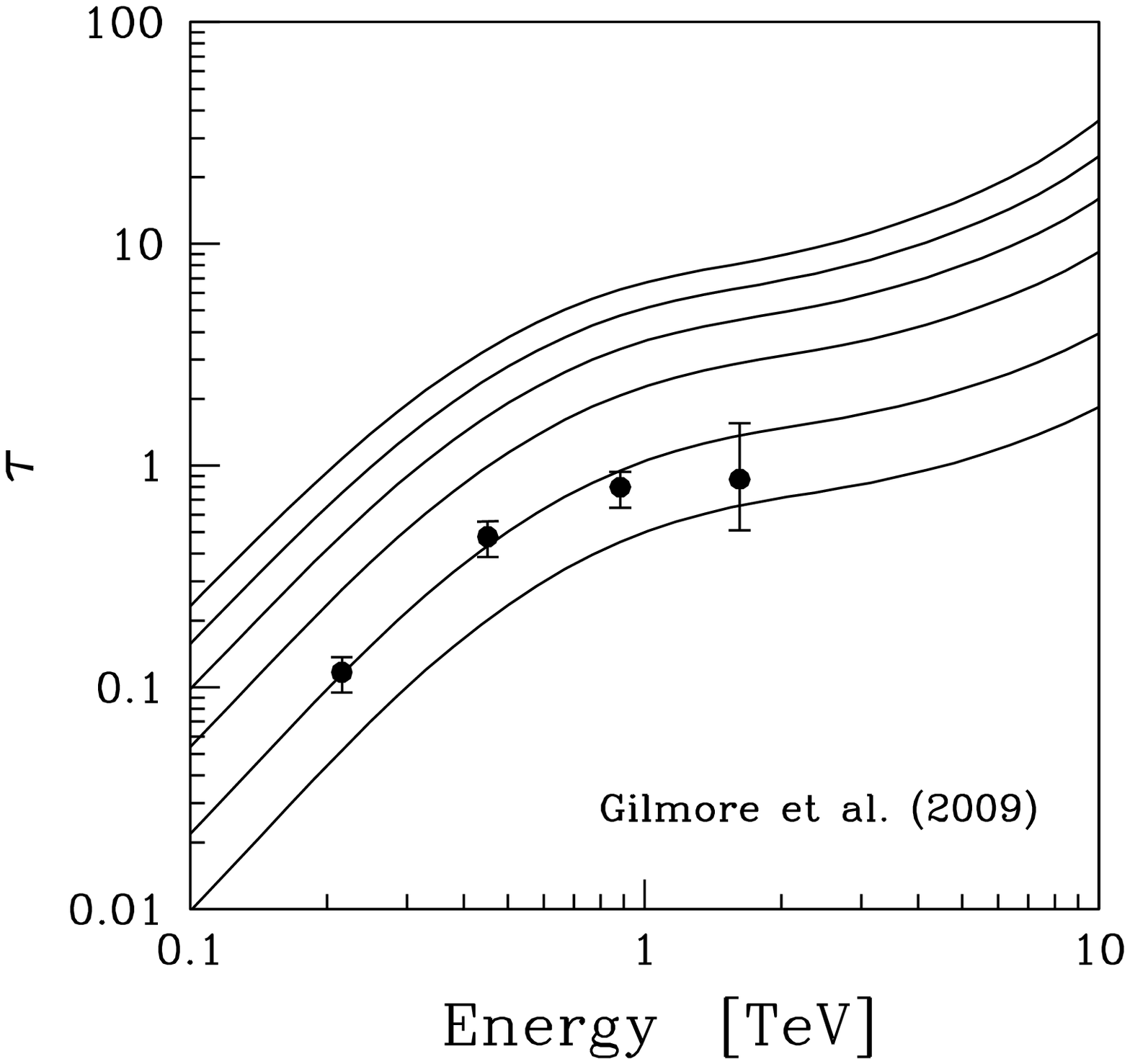}


\vspace{6.5cm}
\includegraphics{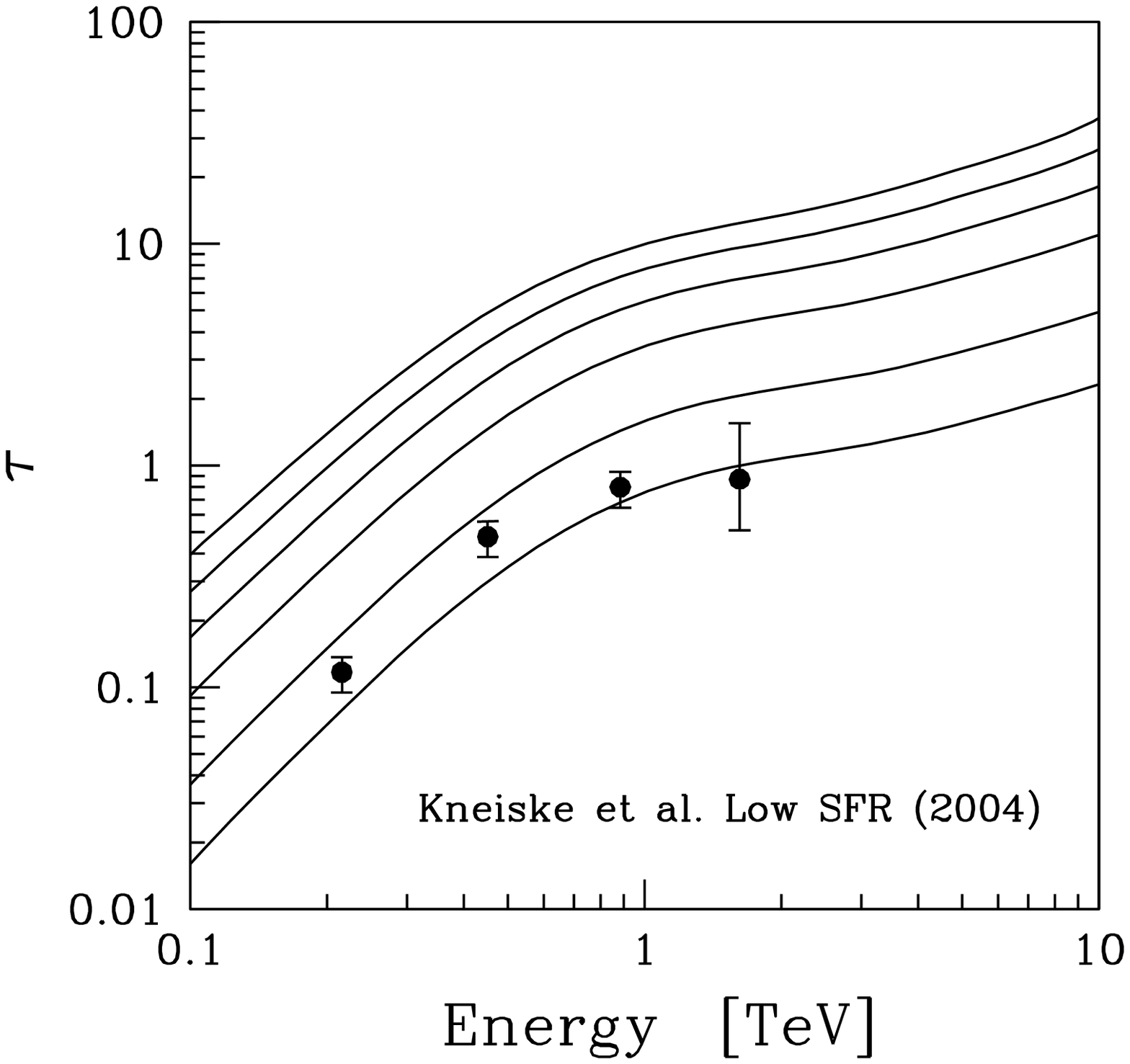}
\includegraphics{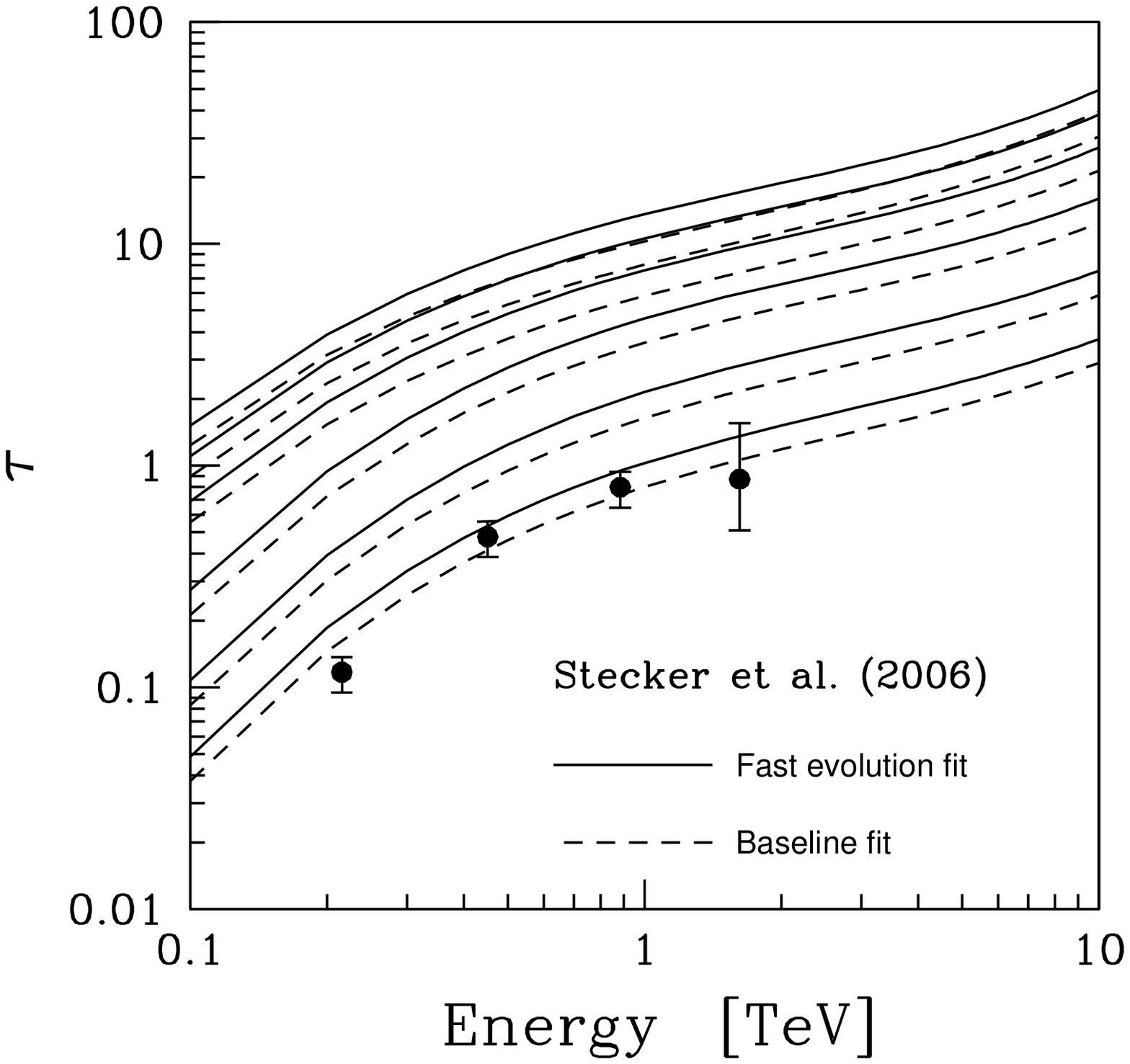}

\vspace{6.5cm}
\includegraphics{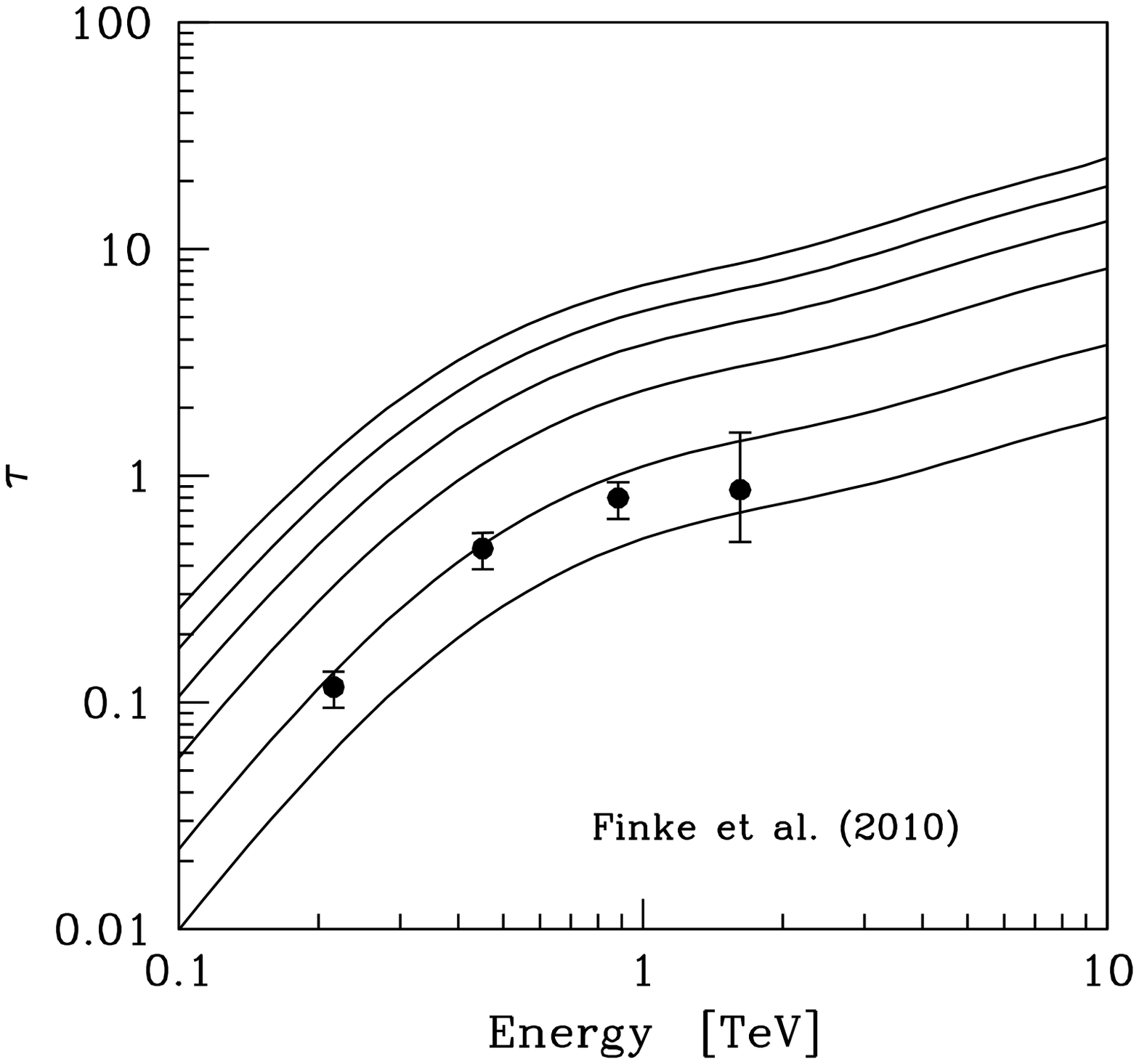}
\includegraphics{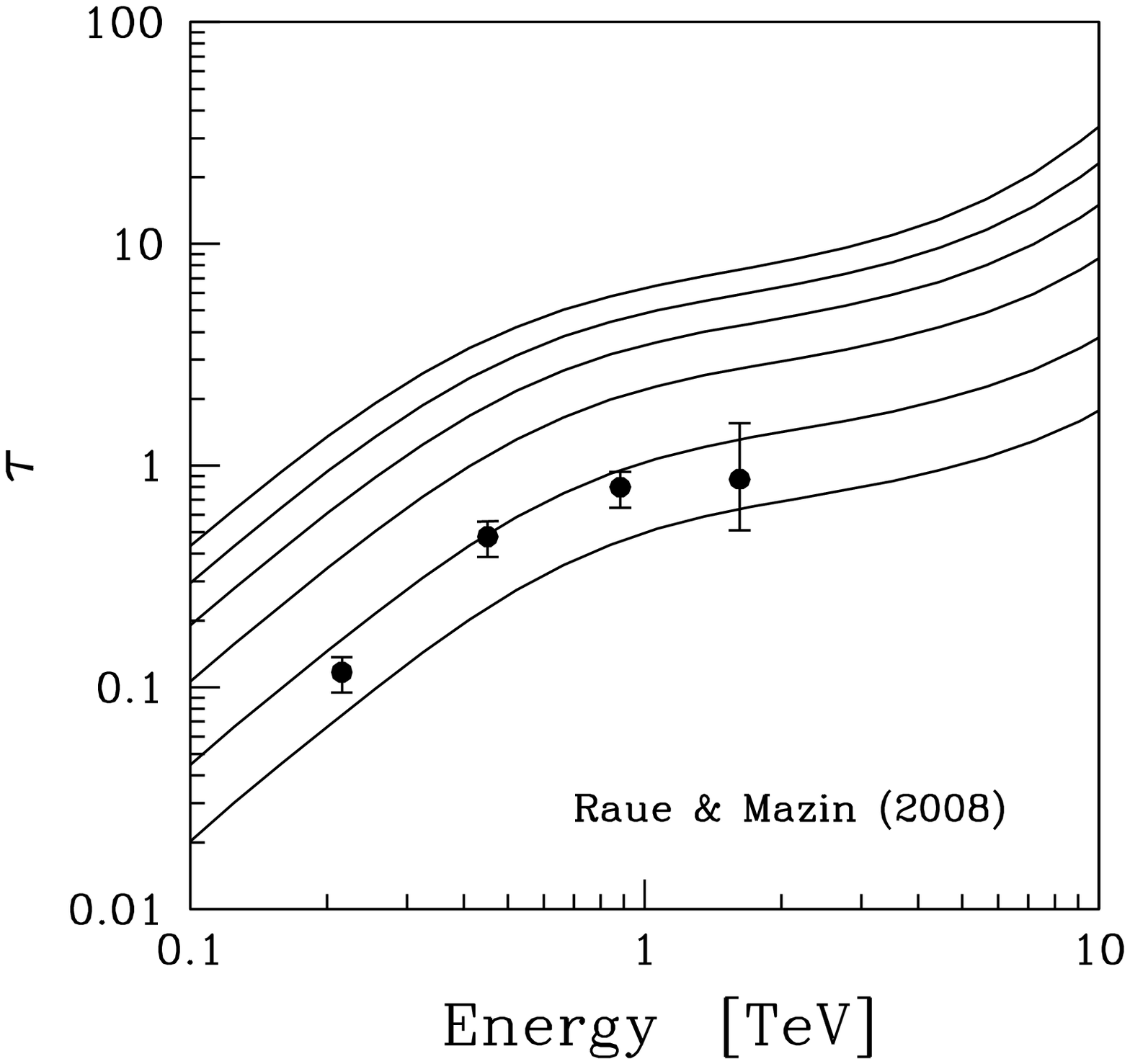}

\end{figure*}

\begin{figure*}

\vspace{6.0cm}

\caption{
Measured values of $\tau_{\gamma\gamma}(E;\,z)$ for $E$$=$0.23, 0.44, 0.88, 
1.70\,TeV derived from comparing, for simultaneous observations of the 
high-frequency-peaked BL\,Lac object 
PKS\,2155-304 ($z=0.116$), the (EBL-affected) VHE\,$\gamma$-ray data 
with the eV-through-GeV best-fitting SSC model extrapolated into the TeV 
domain. Error bars on $\tau$ reflect those on H.E.S.S. data. 
The curves represent, for redshifts $z=0.05$, 0.1, 0.2, 0.3, 0.4, 
0.5 (from bottom up), the optical depth $\tau_{\gamma\gamma}(E)$ 
according to recent EBL calculations: a determination from galaxy counts 
(Franceschini et al. 2008: {\it top left}), two predictions based on a
semi-analytic model of galaxy formation (Gilmore et al. 2009: {\it top 
right}) and a semi-empirical model of the EBL (Kneiske et al. 2004: 
{\it middle left}), a couple of predictions 
based on a galaxy formation model that includes two different galaxy evolution 
rates (Stecker et al. 2006: {\it middle right}), a galaxy formation model that 
takes into account the star formation rate, the stellar initial mass function, 
dust extinction, dust absorption and reradiation, main-sequence and 
post-main-sequence stars (Finke et al. 2010: {\it bottom left}), and a generic 
EBL density (not a complete model) that complies with all existing limits from 
direct and indirect methods (Raue \& Mazin 2008: {\it bottom right}).     }
\end{figure*}

\section{Conclusion} 

The method for measuring the EBL we have proposed in this paper is admittedly 
model-dependent. However, its only requirement is that all the sources used as 
background beamlights should have one same emission model. In the application 
proposed here, we have used a one-zone SSC model where the electron spectrum 
was a (smoothed) double power law applied to the SED of the HBL object 
PKS\,2155-304. While this choice was encouraged by the current observational 
evidence that HBLs seem to have, with no exception, single-slope {\it Fermi}/LAT 
spectra, we could have as well adopted the choice (Aharonian et al. 2009) of a 
triple power law electron spectrum in our search for the best-fit SSC model of 
PKS\,2155-304's SED. Should the latter electron distribution, or any other (e.g., 
curved) distribution, be shown to generally provide a better fit to high-quality 
{\it Fermi}/LAT spectra of HBLs, then that would become our choice. In general, 
what matters to the application of this method, is that {\it all} source SEDs 
be fit with one same SSC model.

Another assumption implicit in our method is that there is an absolute minimum 
in the $\chi^2$ manifold of BL\,Lac emission modeling, and that our 
$\chi^2$-minimization procedure is actually able to find it. Had that not been 
the case for PKS\,2155-304, we would have checked whether the derived 
$\tau_{\gamma\gamma}$s are appreciably different for different model fits.

\acknowledgements 

The work of F.T. was partly supported by an Italian 2007 COFIN-MiUR grant.
We thank Oscar Blanch Bigas, Luigi Costamante, Justin Finke, Laura Maraschi, 
Daniel Mazin, and Floyd Stecker for useful exchanges. 

\bigskip
\bigskip
\bigskip
\bigskip
\bigskip
\bigskip
\bigskip
\bigskip
\bigskip

\def\ref{\par\noindent\hangindent 20pt}

\noindent
{\bf References}
\vglue 0.2truecm

\ref{ Abdo, A.A., et al. (LAT Collaboration) 2009, ApJ, 707, 1310 }
\ref{ Aharonian, F., et al. (H.E.S.S. collab.) 2009, ApJ, 696, L150 }
\ref{ Aharonian, F., et al. (H.E.S.S. collab.) 2006, Nature, 440, 1018 }
\ref{ Aharonian, F.A., Khangulyan, D., \& Costamante, L. 2008, MNRAS, 387, 1206 }
\ref{ Anderhub, H., et al. (MAGIC collab.) 2009, ApJ, 705, 1624 }
\ref{ Blandford, R., \& Eichler, D. 1987, Phys. Rep., 154, 1 }
\ref{ B\"ottcher, M. Dermer, C.D., \& Finke, J.D., 2008, ApJ, 679, L9  }
\ref{ Coppi, P.S., \& Aharonian, F.A. 1999, ApJ, 521, L33 }
\ref{ Costamante, L. (H.E.S.S. collab.) 2008, Intl. J. Mod. Phys. D, 17, 1449 }
\ref{ Costamante, L., \& Ghisellini, G. 2002, A\&A, 384, 56 }
\ref{ Dole, H., et al. 2006, A\&A, 451, 417 }
\ref{ Finke, J.D., Dermer, C.D., \& B\"ottcher, M. 2008, ApJ, 686, 181  }
\ref{ Finke, J.D., Razzaque, S., \& Dermer, C.D. 2010, ApJ, 712, 238 }
\ref{ Finke, J.D., \& Razzaque, S. 2009, ApJ, 698, 1761 }
\ref{ Foschini, L., et al. 2007, ApJ, 657, L81 }
\ref{ Fossati, G., et al. 2008, ApJ, 677, 906 }
\ref{ Franceschini, A., Rodighiero, G., \& Vaccari, M. 2008, A\&A, 487, 837  }
\ref{ Georganopoulos, M., Finke, J., \& Reyes, L. 2010, ApJ, in press (arXiv:1004.0017) }
\ref{ Ghisellini, G., et al. 1998, MNRAS, 301, 451  }
\ref{ Gilmore, R.C., Madau, P., Primack, J.R., Somerville, R.S., \& Haardt, F. 2009, MNRAS, 399, 1694 }
\ref{ Inoue, K., \& Totani, T. 2009, ApJ, 702, 523 }
\ref{ Katarzynski, K., Ghisellini, G., Tavecchio, F., Gracia, J., \& Maraschi, L. 2006, MNRAS, 368, L52 }
\ref{ Kneiske, T.M., \& Mannheim, K., 2008, A\&A, 479, 41 }
\ref{ Kneiske, T.M., Mannheim, K., \& Hartmann, D.H. 2002, A\&A, 386, 1  }
\ref{ Kneiske, T.M., Bretz, T., Mannheim, K., \& Hartmann, D.H. 2004, A\&A, 413, 807  }
\ref{ Krawczynski, H., et al. 2004, ApJ, 601, 151 }
\ref{ Kusunose, M., \& Takahara, F. 2008, ApJ, 682, 784 }
\ref{ Malkan, M.A., \& Stecker, F.W. 1998, ApJ, 496, 13 }
\ref{ Malkan, M.A., \& Stecker, F.W. 2001, ApJ, 555, 641 }
\ref{ Mazin, D., \& Goebel, F. 2007, ApJ, 655, L13 }
\ref{ Mazin, D., \& Raue, M. 2007, A\&A, 471, 439  }
\ref{ M\"ucke, A., Protheroe, M.J., Engel, R., Rachen, J.P., \& Stanev, T. 2003, Astropart. Phys., 18, 593 }
\ref{ Raue, M., Kneiske, T., \& Mazin, D. 2009, A\&A, 498, 25  }
\ref{ Raue, M., \& Mazin, D. 2008, Int.J.Mod.Phys.D17, 1515 }
\ref{ Razzaque, S., Dermer, C.D., \& Finke, J.D. 2009, ApJ, 697, 483 }
\ref{ Rephaeli, Y. 1979, ApJ, 227, 364 }
\ref{ Salamon, M.H., \& Stecker, F.W. 1998, ApJ, 493, 547  }
\ref{ Schroedter, M. 2005, ApJ, 628, 617 }
\ref{ Stecker, F.W., Baring, M.G., \& Summerlin, E.J. 2007, ApJ, 667, L29  }
\ref{ Stecker, F.W., \& de\,Jager, O. 1998, A\&A, 334, L85  }
\ref{ Stecker, F.W., \& de\,Jager, O.C., \& Salamon, M.H. 1992, ApJ, 390, L49 }
\ref{ Stecker, F.W., Malkan, M.A. \& Scully, S.T. 2006, ApJ, 648, 774 }
\ref{ Stecker, F.W., \& Scully, S.T. 2010, ApJ, 709, L124  }
\ref{ Tagliaferri, G., et al. (MAGIC collab.) 2008, ApJ, 679, 1029 }
\ref{ Tavecchio, F., \& Maraschi, L. 2003, ApJ, 593, 667 }
\ref{ Tavecchio, F., Maraschi, L., \& Ghisellini, G. 1998, ApJ, 509, 608  }
\ref{ Tavecchio, F., Ghisellini, G., Ghirlanda, G., Foschini, L., \& Maraschi, L. 2010, MNRAS, 401, 1570 }
\ref{ Tavecchio, F., et al. 2001, ApJ, 554, 725 }
\ref{ Venters, T.M. 2010, ApJ, 710, 1530 }
\ref{ Venters, T.M., Pavlidou, V., \& Reyes, L. 2009, ApJ, 703, 1939 }

\end{document}